\documentclass{sig-alternate-yuyinben}

\usepackage{algorithm}
\usepackage{algorithmic}

\begin{document}

\title{A Method to Support Difficult Re-finding Tasks}

\numberofauthors{2}
\author{
\alignauthor
Gangli Liu\\
       \affaddr{Department of Computer Science}\\
       \affaddr{Tsinghua University}\\
       \email{gl-liu13@mails.tsinghua.edu.cn}
       \alignauthor
Ling Feng\\
       \affaddr{Department of Computer Science}\\
       \affaddr{Tsinghua University}\\
       \email{fengling@mail.tsinghua.edu.cn}
}

\maketitle
\begin{abstract}
Re-finding electronic documents from a personal computer is a frequent demand to users. In a simple re-finding task, people can use many methods to retrieve a document, such as navigating directly to the document's folder, searching with a desktop search engine, or checking the Recent Files List.

However, when encountering a difficult re-finding task, people usually cannot remember the attributes used by conventional re-finding methods, such as file path, file name, keywords etc., the re-finding would fail.

We propose a new method to support difficult re-finding tasks. When a user is reading a document, we collect all kinds of possible memory pieces of the user about the document, such as number of pages, number of images, number of math formulas, cumulative reading time, reading frequency, printing experiences etc. If the user wants to re-find a document later, we use these collected attributes to filter out the target document.

To alleviate the user's cognitive burden, we use a question and answer wizard interface and provide recommendations to the answers for the user, the recommendations are generated by analyzing the collected attributes of each document and the user's experiences about them.
\end{abstract}

\keywords{Re-finding; Personal Information Management; Information Retrieval}

\section{Introduction}
As computers are becoming an indispensable tool for people, re-finding a document which has been accessed previously becomes a common task \cite{Jensen2010}.  Some of the re-finding tasks are effortless to accomplish, especially when the user is familiar with the target file or the document has been accessed recently. For a simple re-finding task, the user can navigate directly to the folder which contains the target document and access it. It is an instinctive behavior of people, and is the most common way to accomplish re-finding tasks \cite{Capra2003}. If a user cannot recall a document's path but remembers its keywords or other metadata attributes such as size, creation time, author etc., a desktop search engine can be used to retrieve the document.\\

Sometimes, the user's memory about the target document is very vague, and cannot recollect effective attributes (path, keywords, metadata etc.) to filter out the target, the re-finding task becomes a difficult one. In our opinion, under these conditions, maybe a user can recall some other attributes which are not used by present file retrieval systems. The user may remember some other knowledge or experiences about the target document, which have not been exploited. For example, suppose a user is re-finding a document, she cannot remember its path and keywords, but she clearly remembers the document has about 10 pages, contains at least 2 images, she has only read the abstract part of the document, and has printed it sometime. These information can be useful, but have not been collected and exploited by present file retrieval systems.\\

Sometimes a user may recall many attributes of a document, but none of them is specific enough to locate the file, all of them are general impressions, this forms a difficult re-finding task too, because none of them can provide an effective clue to retrieve the document. We can use group effort to achieve the re-finding task. For example, a user may not remember a document's specific path, but she is sure it is in E drive (for Microsoft windows operating system); she may not remember its specific size, but she knows it is above 1M; she may not remember its specific last access date, but she clearly remembers it's a weekend. If each of these general impression information can exclude 1/3 of unrelated documents, 16 attributes can filter a candidate set of 10,000 into about 15. Then, we can use just scan and recognition to get the target document. If a user can recall a more specific value about a particular attribute, fewer filter attributes can be used.

\subsection{How people's memory works\cite{Mohs2007}}
When we are reading a document, memory encoding is the first step to create our memory, psychologists believe there are two parts of the brain, the hippocampus and the front cortex, are responsible for analyzing various sensory inputs and decide if they are worth remembering.\\

People's memory can be stored in three stage: the sensory stage, the short-term stage, and ultimately, the long-term stage. Each stage of human memory can function like a filter which discards useless information. The sensory memory can only last a fraction of a second, the short-term memory can only hold about seven items for 20 to 30 seconds at a time. Only those important information can be gradually transferred into long-term memory.

\subsection{Record all the possible attributes}

When people want to re-find a document, only those attributes stored in long-term memory can be used to locate the target. A document can have many attributes, such as file name, size, keywords, and authors etc. A user's experiences about a document can also be used to locate the target, such as printing the document, reading the document at a special place or time etc.
Therefore, a lot of attributes can be used to re-find a document, as long as two conditions are satisfied: Firstly, the user has a long-term memory about the attributes. Secondly, an information retrieval system has recorded the attributes' values for all the documents.\\

There is uncertainty which set of attributes will get into a user's long-term memory. It is decided by various factors, such as the user's interest, characteristics, experiences, and other circumstance factors.
We propose using a logging system to record all the possible attributes when the user is processing a document, in case anyone of the recorded attributes would be used as a filter to locate the target document. These attributes can include a document's intrinsic attributes and the user's experiences about the document. The quantity of documents a user can access during their life is limited, so the total recorded information will not be large.

\subsection{Alleviate the user's cognitive burden}

Demanding a user to recollect many kinds of information about a file will bring the user unacceptable cognitive burden. According to the findings of cognitive psychology, when people are trying to recollect something, a great deal of their memory is \textit{``available but not accessible"} \cite{Tulving1966,Tulving1973}. For example, if you ask a user to provide all the information about a document she has read, she may give you little information about it, because a lot of information are \textit{``available but not accessible"}. Therefore, we use a question and answer wizard interface to collect a user's memory about the target document.\\

 A logging system is used to record all the possible information about each document and the user's reading experiences of the document. Some information cannot be extract directly, such as how long and how much the user has read the document, an analysis system is also needed.\\

To further alleviate the user's cognitive burden, we try to give recommendations for each question of the list, the recommendations are generated based on analyzing the database which contains each file's attributes and the user's experiences about each file. As mentioned above, in a difficult re-finding task, a user's memory about the target file usually is some general impressions, so our recommendations will be generalized to a general impression. For example, suppose one of the questions asks the user:\\

\textit{``Do you remember how many tables the document contains?"}\\

Instead of recommending concrete numbers like 2, 3 or 5, we will recommend a generalized answer like:\\

\textit{``A. above 3 \quad  B. below or equal 3"}. \\

The specific parameter will be generated base on analyzing the database.\\
Based on our method, we implemented a preliminary system to help users re-find PDF (Portable Document Format) documents. Section 3 introduces the implementation details about the re-finding system.

We also recruited 5 users to test our PDF re-finding system. Section 4 shows the results of the user study.\\
Figure 1 shows the framework of our method.
\begin{figure}
\centering
\includegraphics[height=2.7in, width=3.5in]{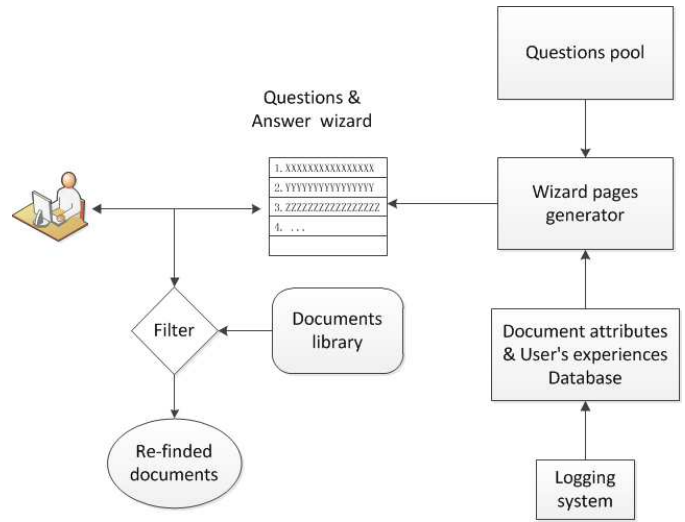}
\caption{The framework of the method.}
\end{figure}

\section{RELATED WORK}
Personal Information Management (PIM) systems help people to acquire, store, organize and re-find information \cite{Jones2007}. Re-finding is a significant part of a PIM system, without re-finding, there is no need to store and organize information.\\

Re-finding is a frequent activity in people's daily lives. McKenzie and Cockburn show 60-80\% of web page visitations are re-access \cite{McKenzie2001}. 40\% of web searches are to re-find a piece of information which the user has seen before \cite{Teevan2007, Tyler2010}. Re-visitation of emails is less common compared to web pages and documents, Elsweiler et al. showed 6-15\% of emails will be re-founded later \cite{Elsweiler2011}.\\

There are two psychological processes involved in a re-finding task, recall-directed search and recognition-based scanning \cite{Lansdale1988}. People pay careful attention when naming a file, primarily not for using the name to search, but for recognizing the file when it appears in a list of candidates \cite{Barreau1995a,Nardi1995}. In a desktop search environment, file names are frequently used in the recall-directed search process too.\\

Bergman et al. note navigating through a file hierarchy is the most common method people employ to re-find their files \cite{Bergman2008}. Navigation has a significant advantage, it can feedback some cues while the user is navigating between file folders, these cues may help retrieve more memories about the target files, like file name keywords, or their relationship with other files. Besides, navigation makes scan and recognition behavior easier because files contained in the same folder usually share some common characteristics, such as they belong to same project or they are different versions of a same document. These advantages make navigation the most popular method used by people to re-find files. Azagury et al. propose a similar navigation paradigm named \textit{``faceted search"}, allows users to navigate based on file attribute values, such as file type and modification time \cite{Azagury2002}.\\

However, when the user's memory about the target document is misty or erroneous, s/he will get \textit{`lost'} in the folder hierarchies, viewing same folders or even same files multiple times \cite{Elsweiler2011}.

Deng et al. devised a \textit{`ReFinder'} system which utilizes the user's previous access context (such as location, access time, background computer activities etc.) to re-find a web page \cite{Deng2013}. It takes memory degradation into consideration and tackle it by using a corresponding degrading context tree.\\

Qvarfordt et al. designed a \textit{`SearchPanel'} Chrome browser extension which records a user's process history information of the search result page (SERP), provides sense-making, navigation, and re-finding functions based on the recorded information during a complex information need \cite{Qvarfordt2014}. Chau et al. developed a \textit{`Feldspar'} system which can retrieve personal files by associations related to other objects, such as email, person, folder or event etc. They also devised an algorithm collecting associations between objects \cite{Chau2008}.\\

Fitchett and Cockburn designed an algorithm, \textit{`AccessRank'}, which predicts a user's re-visitation targets, it uses recency, frequency, temporal clustering and time of day to generate its predictions \cite{Fitchett2012}.

\textit{``Stuff I've Seen"} is a system which help users re-find their previously accessed documents, it integrates web pages, emails and documents into a centralized system, illustrates that time and associated people are important retrieval cues to assist re-finding \cite{Dumais2003}. A further study shows that segment the timeline using landmark events experienced by the user will decrease re-finding time \cite{Ringel2003}. People usually have poor memory about abstract dates, if associated with their personal events, the recalled information inclines to be clear and accurate.\\

Boardman and Sasse found people like to organize their files \cite{Boardman2004}, the main purpose is for easy re-finding \cite{Barreau1995}. 97\% of their files will be placed in an intended folder, only 3\% of their files will be left in some default locations. Accordingly, people usually have some impression about their file hierarchy, even if it may be obscure. When they cannot designate the target file's precise location, perhaps they can specify a list of candidate folders which may contain the document.\\

Bergman et al. found users are inclined to organize their personal files by projects (e.g., all the files about an algorithm design course), rather than file type \cite{Bergman2006}. This implies we can use folders to deduce a user's projects and establish an association between projects and documents

\section{A PDF DOCUMENT RE-FINDING SYSTEM}
To verify our method, we developed a PDF document re-finding system, which helps a user re-find PDF documents she has read. PDF documents are very common for knowledge workers like researchers. It is common for knowledge workers to re-find PDF documents like papers or electronic books.\\

The re-finding system is mainly composed of two parts: the analysis and logging subsystem and the re-finding wizard subsystem. The analysis and logging subsystem records a pdf document's various attributes and the user's reading experiences about the document. The re-finding wizard subsystem provides an interface to fulfill the user's document re-finding demands.
\subsection{The analysis and logging subsystem}
To implement the analysis and logging subsystem, we developed a plug-in to the Adobe Acrobat reader application. At present the plug-in only supports Microsoft Windows operating system, other platforms will be supported in future. When the user is reading a PDF document, the plug-in records all the necessary data into a database.
We classify the attributes being collected into two categories: the intrinsic document attributes and the user's experiences about a document.
\subsubsection{The intrinsic document attributes}
The intrinsic document attributes are those attributes holding by a document constantly, it will not change when the user operates a document in read-only mode, such as number of pages the document has, number of images, file size etc.
If a document is edited by the user, some of the intrinsic document attributes may change. In our implementation, we differentiate the edited document as another file from the original document. In practice, people seldom edit a PDF document.
Table 1 shows the intrinsic document attributes we are collecting now, new items can be added in future.

\begin{table*}
\centering
\caption{The Intrinsic Document Attributes}
\begin{tabular}{|c|c|c|} \hline
Attribute  name&Abbreviation&Description\\ \hline
File path & FP & \\ \hline
File name & FN & \\ \hline

File creation date& 	FCD&	The date the author produced the document \\ \hline

File acquisition date&	FAD	&The date the user obtained the document \\ \hline
Pages	& &	Number of Pages the document contains \\ \hline

Number of images &	Images	& \\ \hline
Number of tables	&Tables	& \\ \hline
Number of math formulas&	Formulas&	\\ \hline
Bibliography& &		Does it contain a bibliography? \\ \hline

\hline\end{tabular}
\end{table*}

\subsubsection{User experience attributes}
User experience attributes are those attributes which involve the user's process behaviors about a document.
Table 2 shows the user experience attributes we are collecting now, new items can be added in future if available.

\begin{table*}
\centering
\caption{The User Experience Attributes}
\begin{tabular}{|c|c|c|} \hline
Name	&Abbreviation	&Description\\ \hline
Printing experience&	PE&	Whether printed it before?\\ \hline
Re-finding experience&	RE	&Whether retrieved it before?\\ \hline
Unusual access time	& UAT&	Whether accessed it at night or weekend?\\ \hline
Unusual access location	&UAL	&Whether accessed it at an unusual place?\\ \hline
Access frequency	& AF&	How many times has the user accessed it?\\ \hline
Accumulative process time&	APC	 &\\ \hline
Coverage percentage	&Coverage&	Have the user read the full document or only a part?\\ \hline
Last access time&	LAT	&\\

\hline\end{tabular}
\end{table*}

\subsubsection{An algorithm to calculate a user's cumulative process time on a document}
To calculate a user's precise process time on a document, various conditions should be considered.\\

\textbf{1. The user opens a document does not mean the user is reading the document.}\\

It is normal that a user is reading a document while doing other things, such as programming. So simply using the document's close date minus open date is not a good idea, this strategy will bring in invalid document processing time.
When the user starts the Adobe Acrobat reader application, our plug-in starts a thread which periodically checks whether the reader application is the foreground window. If the reader application is not the foreground window, the timer stops.\\

\textbf{2. The reader application is the foreground window does not mean the user is reading a document.}\\

It is a normal situation that the user is reading a document and then leaves for some reasons, with the document still opening.
To solve this problem, our plug-in starts an independent thread to check the user's mouse and keyboard input patterns, if there is no input within a period of time T0, we think the user has left, the timer stops.
In section 4 we will discuss how to decide the parameter T0, by analyzing the user's mouse and keyboard inputs interval data.\\

\textbf{3. The user may have several copies or several versions of the same document.}\\

It is common that a user copies a document to several folders, reading each of them for a period of time. Although they are different documents physically, but when the user reads them, she do not care which copy she is reading, in her mind, they are the same document. Therefore, when we update a document's cumulative reading time, we also update other copies' or versions' cumulative reading time.

Algorithm 1 shows how we calculate a user's cumulative process time on a document.

\begin{algorithm}
\caption{An algorithm to calculate a user's precise process time on a document.}
\label{alg1}
\begin{algorithmic}

\WHILE{ReaderAPPisRunning}

\STATE	ct1 = GetCurrentTime();
\IF{Reader APP is the foreground window}

\IF{Input Interval < Max Input Interval T0}
\STATE // T0 is decided in section 4
\STATE CurrentDoc = AppGetCurrentDoc();

\IF{CurrentDoc == LastDoc}

\STATE DocProcessTimeAdd(CurrentDoc,T);
\STATE CurrentPage = AppGetCurrentPage();

\IF{CurrentPage == LastPage}

\STATE PageProcessTimeAdd(CurrentPage,T);
\ELSE
\STATE LastPage = CurrentPage;
\ENDIF
\ELSE
\STATE LastDoc = CurrentDoc;
\ENDIF
\ENDIF
\ENDIF
\STATE ct2 = GetCurrentTime ();
\STATE timeLoss = $ct2 \quad $-$ \quad ct1$;
\STATE Sleep($T - timeLoss$); // currently we set T = 3 seconds
\ENDWHILE

\end{algorithmic}

\end{algorithm}

A thread periodically checks the current active document and its active page, if the active document is the same with the last one, we add a time period to its cumulative process time, a page's cumulative process time is calculated similarly.
Because database operations are relatively time-consuming jobs, we set a counter to record the time loss brought by database operations, and then compensate the time loss in the next loop.

\subsubsection{Document process coverage}
The user usually has a clear memory of how much she has read the document. For example, when a user is reading a scientific paper, she does not have enough time or interest to read it completely, and just read the abstract part of the document. This information can be used to help the user re-find the document.

As introduced in section 3.1.3, we record a user's cumulative reading time of each page of a document. With this information, it is easy to calculate a user's coverage percentage of a document: sum all the pages the user has read, then divide it by the total pages the document contains.\\
Recording a user's reading time of each page can also help the user quickly locate the wanted information. Suppose the user has already retrieved a document, by the recorded page time, we can learn which pages the user has read most, and bring the user to those pages quickly.

\subsubsection{Access frequency}
Usually the user cannot remember exactly how many times she has accessed the document, but if our question is about a general impression, like less than 3 times or more than 3 times, the user usually can provide a correct answer. So we record a user's access frequency about a document as a re-finding attribute.
There is a question needs to consider. If a user opens a document and then finds this is not what she wants, so she closes the document. This access experience should not be considered as an effective one. We set a timer to record a document's open date and close date, if their interval is less than a particular threshold, like 10 seconds, the access frequency attribute will not be updated.

\subsubsection{The re-finding wizard subsystem}

The re-finding wizard subsystem provides an interface to collect the user's memory pieces about the target document, then shows the re-finding results to the user.
At each step of the wizard, we ask the user a question about the correspondent attribute, such as: \textit{``How many pages the document has?"}\\
To alleviate the user's cognitive burden, we provide recommendations to the answers, such as:\\

\textit{``A. More than 10 pages  B. Less than or equal to 10 pages"}\\

The parameters of the recommendations are calculated by analyzing the distribution of the attribute values. The analysis and logging subsystem has recorded these attribute values for each document when the user reads it.
We provide an option for the user to give a more precise answer than our recommendations. It is possible that the user has a very clear memory about a particular attribute due to some special reasons. For example, the user noticed a document contains two images of two girls, one looks very charming and the other looks ugly. This dramatic contrast effect attracts her attention, so she remembers this attribute clearly. Later she wants to re-find the document, we ask a question of \textit{``How many images does the document contain?"}, and she can give us a precise answer of 2. With this precise information, it is easy to locate the target document.\\

For a difficult re-finding task, it is common that the user is not sure about her answer to a question. We provide three options to let the user describe how confident she is about her answer:\\

1.\textit{``I'm sure!"}
2.\textit{`` I'm not sure."}
3.\textit{`` I don't know the answer."}\\

After the answer collecting phrase, we first use the attributes on which the user has chosen the first option, if those attributes are not enough to filter out the target document, attributes on which the user has chosen the second option will be used. If the user has chosen the third option, this attribute will be ignored during the re-finding process.
To further help the user obtain the correct answer to a question, we also provide a picture showing the global distribution of the attribute values as a reference. In section 4 we will evaluate whether this picture is helpful for re-finding.\\
Figure 2 and Figure 3 show two example pages of the wizard during a re-finding process.
\begin{figure}
\centering
\includegraphics[height=2.55in, width=3.5in]{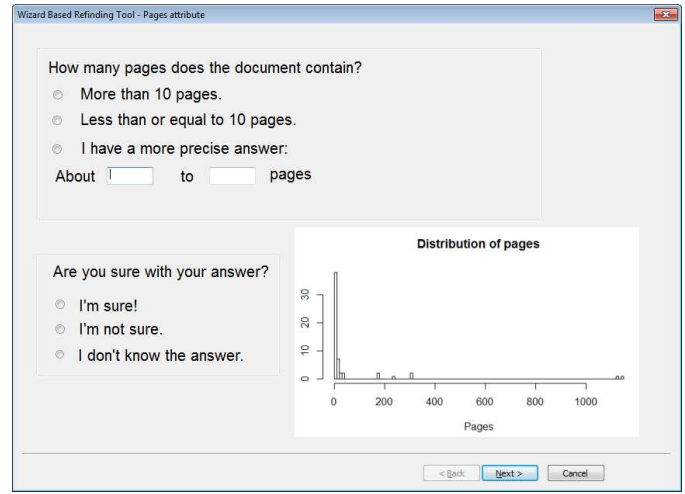}
\caption{The Pages attribute.}
\end{figure}

\begin{figure}
\centering
\includegraphics[height=2.55in, width=3.5in]{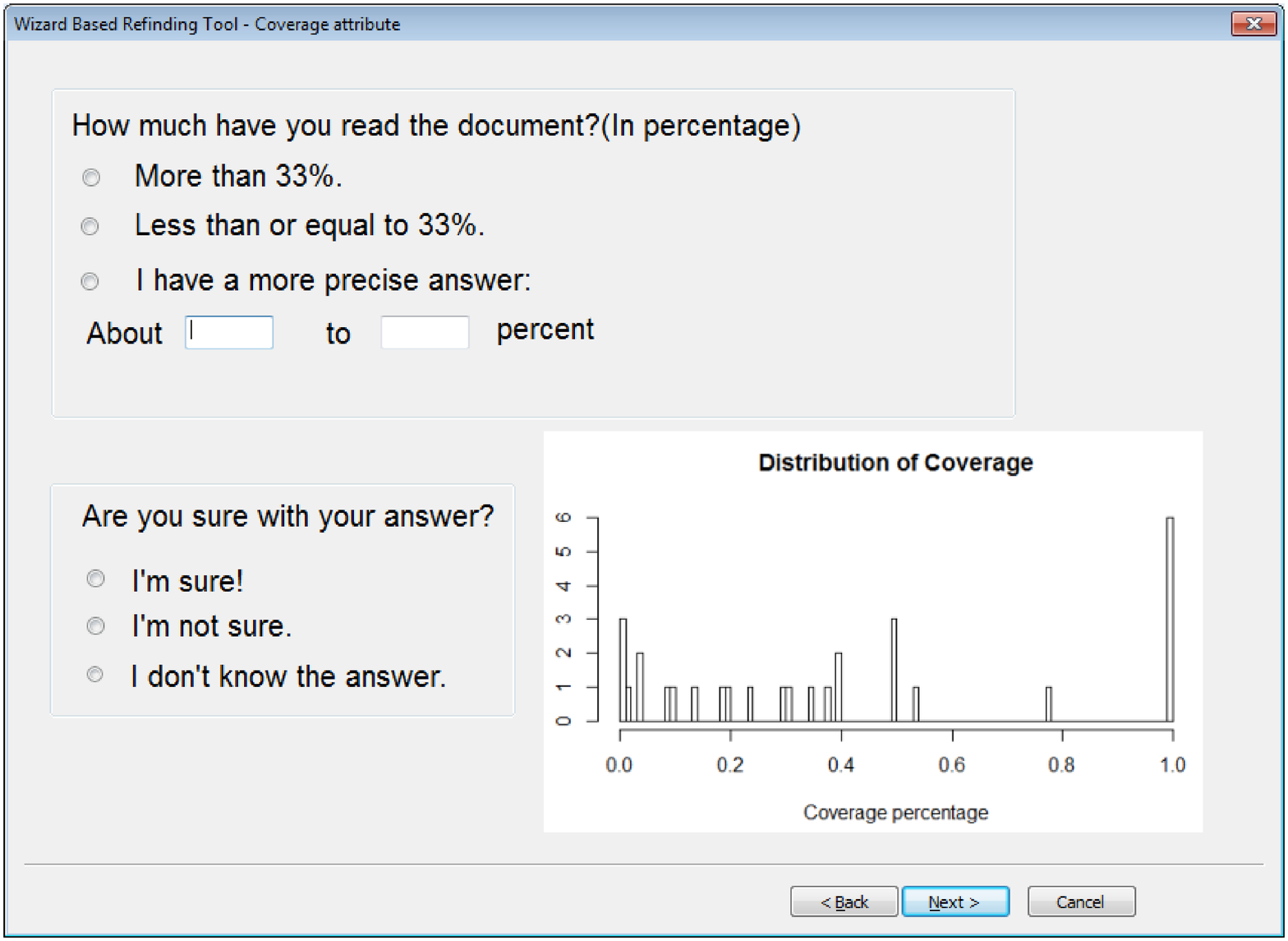}
\caption{The Coverage attribute.}
\end{figure}

Figure 4 shows the re-finded results, to keep the neatness of the interface, at present we only show 6 attributes of the document, that is: document ID, path, number of pages, total process time(in seconds), access frequency, and coverage percentage. Other attributes can be added if needed.

\begin{figure}
\centering
\includegraphics[height=2.08in, width=3.5in]{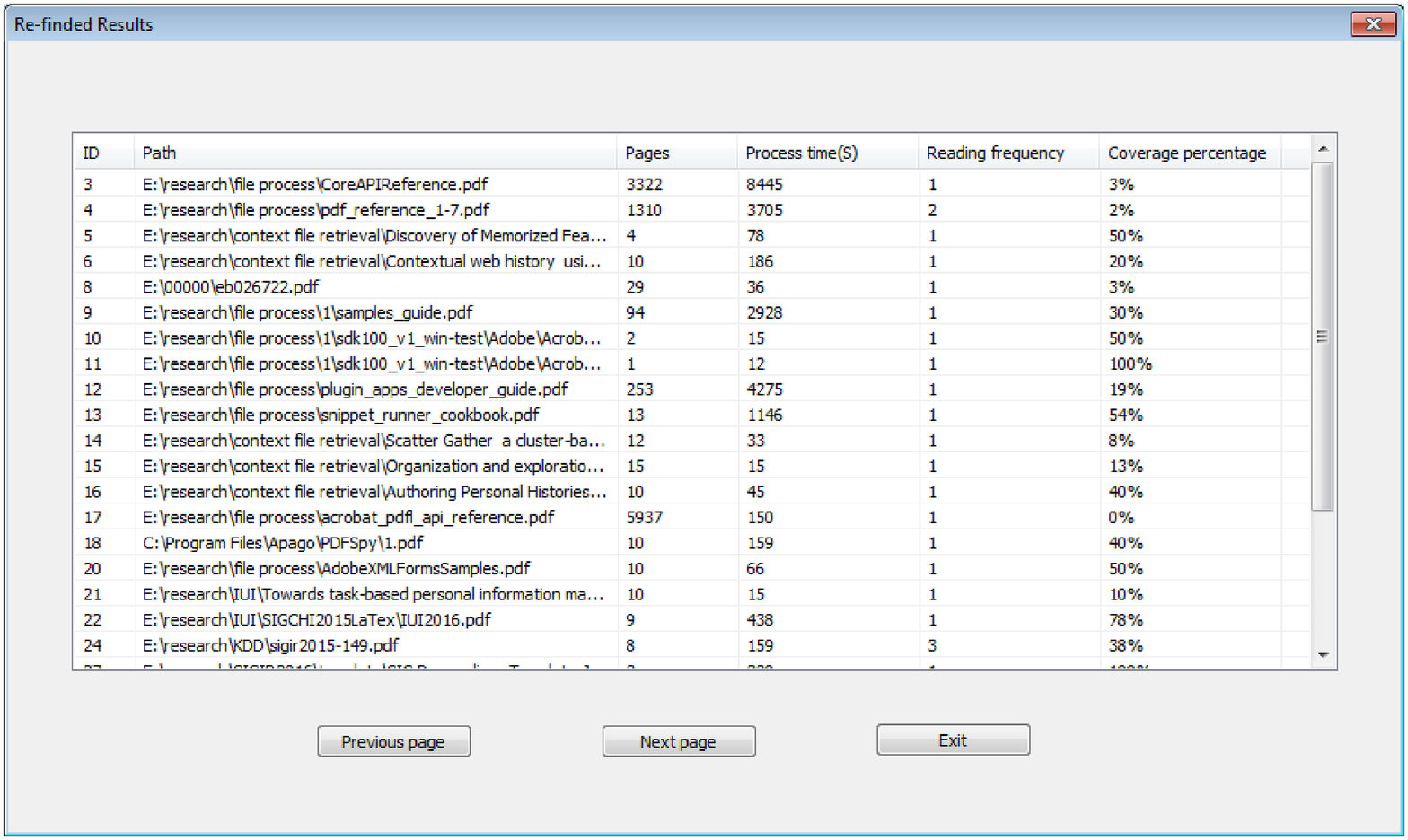}
\caption{The re-finded results.}
\end{figure}

\section{USER STUDY}
To evaluate how our method can perform, we recruited 5 users to test our PDF re-finding system. All of them are research students, 4 male and 1 female, aged from 22 to 34.
In addition, we recorded 4 users' mouse and keyboard input intervals when they are reading PDF documents. By analyzing these recorded data, we can obtain an optimal value for the parameter T0 mentioned in Algorithm 1 of section 3.1.3.

\subsection{Some statistics of the users' reading behavior}

Table 3 shows some statistics of the data recorded by the analysis and logging subsystem.

\begin{table*}
\centering
\caption{Some Statistics of The Recorded Data}
\begin{tabular}{|c|c|c|c|c|c|c|c|} \hline

	&Time range	&Total PDF &	Total reading&	Average Reading&	Average document &	Average coverage &	Average reading \\

&&numbers& count&  time/Docment(S) &size(Pages)&percentage& frequency\\ \hline

U1	&3 weeks&	26&	37	&1187&	566&	29\%&	1.4\\ \hline
U2	&4 weeks&	132&	236&	836&	29&	36\%&	1.8\\ \hline
U3	&3 weeks&	29	&55	&2280	&15	&18.7\%	&1.9\\ \hline
U4	&3 weeks&	39&	54	&677&	41	&64\%&	1.4\\ \hline
U5	&3 weeks&	45&	68	&2369&	13&	52\%&	1.5\\ \hline

\hline\end{tabular}
\end{table*}

\subsection{The PDF re-finding system's performance}

To test the performance of our PDF re-finding system, we randomly selected some documents and let the user re-find them with our tool. For each document, the user is asked 15 questions listed in Table 1 and Table 2. File path and file name attributes are not evaluated, because under a difficult re-finding context, these two attributes usually will not be remembered by a user.
The results shows that our PDF re-finding system's average success rate is 55\%. The users are asked whether the distribution pictures are helpful during their re-findings, 80\% of users give a positive answer.
Table 4 shows the results of performance evaluation.

\begin{table*}
\centering
\caption{The Results of Performance Evaluation}
\begin{tabular}{|c|c|c|c|c|c|c|}  \hline

& Re-finding&	Questions which have&	Questions which have&	Questions ignored	&Whether The Distribution&Success\\

& tasks& confident answers & uncertain answers& & Picture Helpful & rate  \\ \hline

U1&	10&	73.3\%&	14\%&	12.7\%&	YES&	60\%\\ \hline
U2&	20&	40\%&	33.3\%&	26.7\%&	YES&	45\%\\ \hline
U3&	10&	66\%&	20.7\%&	13.3\%&	YES&	70\%\\ \hline
U4&	10&	58.7\%&	27.3\%&	14\%&	NO&	60\%\\ \hline
U5&	10&	53.3\%&	26\%&	20.7\%&	YES&	40\%\\ \hline

\hline\end{tabular}
\end{table*}

During the performance evaluation, we also recorded users' memory strength about each attribute. Questions that the user has a confident answer implies a good memory of that attribute, questions the user ignored implies a bad memory of the attribute. Based on these standards, we obtained users' memory strength about each attribute. Figure 5 shows each attribute's score. From the scores, we can conclude that people usually have a good memory of whether a document contains math formulas, whether she has printed a document, and the last access time of a document.

\begin{figure}
\centering
\includegraphics[height=1.85in, width=3.5in]{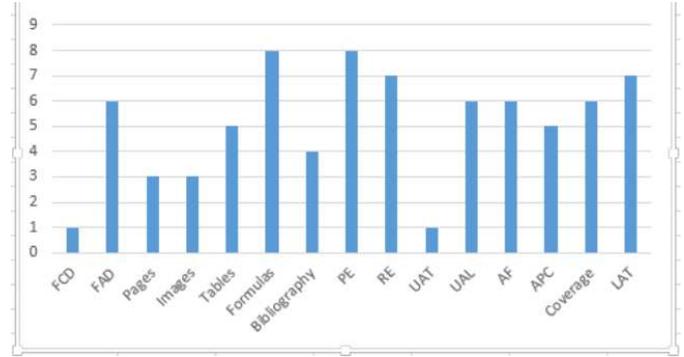}
\caption{User's memory strength about each attribute.}
\end{figure}

\subsection{Analysis of users' mouse and keyboard input interval patterns}

To learn the user's PDF reading habits and obtain the optimal parameter T0 of Algorithm 1 in section 3.1.3, we recorded 4 of the 5 users' input interval information for about 3 weeks.\\

Different activities may result in different input patterns. For example, watching movies and playing games on a computer have totally different mouse and keyboard input patterns.
To eliminate the interference of the difference of user activities, we only record the users' mouse and keyboard input intervals when the PDF reader application is the foreground window. That is to say, we only record the user's input intervals when she is reading a PDF document.\\

The PDF reader application is the foreground window does not mean the user must be reading a document, maybe she has left and left the computer opening. To get the precise value of a user's reading time of a document, we need to know whether the user is still there or has left.
If there is a mouse or keyboard input, surely we can conclude the user it still there, reading the document. If there is no mouse or keyboard input for a long time, such as 1 hour, we can conclude that the user has left, then we should stop the timer of the user's reading time.\\

One question is when we should stop the timer after we cannot detect any user inputs. If we stop too early, maybe the user has not left, she is just thinking hardly of a complicated formula of the document, this will bring in a type I error of the reading time; If we stop too late, maybe the user has left for lunch, but our timer is still counting, this will bring in a type II error of the reading time.
Figure 6 and Figure 7 show two randomly selected users' input interval distributions, for a time rang of about 3 weeks.

\begin{figure}
\centering
\includegraphics[height=2.05in, width=3.5in]{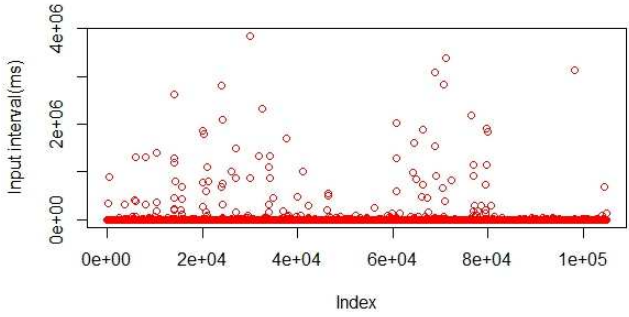}
\caption{User3's input intervals distribution.}
\end{figure}

\begin{figure}
\centering
\includegraphics[height=2.05in, width=3.5in]{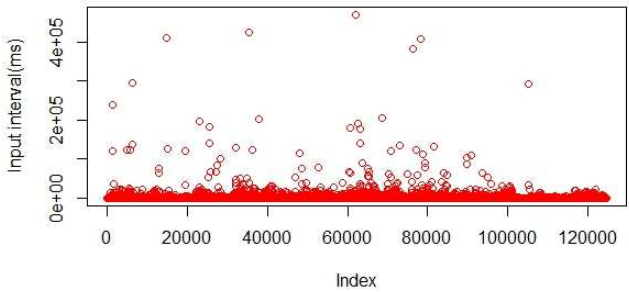}
\caption{User4's input intervals distribution.}
\end{figure}

By analyzing a user's input interval patterns, we can obtain the optimal parameter of when we should stop the timer. Different users may have different document reading habits, the recorded data can help us obtain a personalized parameter.
Table 5 shows the users' input intervals data, we can see that overwhelming majority of the users' input intervals are less than 1 minute. For User1 and User2, if there is no inputs within 20 minutes, that implies the user has left.  For User4, if there is no inputs within 10 minutes, that indicates he has left. User3 has a special reading habit than others, he can read a document for an hour without moving mouse or striking keyboard.

\begin{table*}
\centering
\caption{The Users' Input Intervals Data}
\begin{tabular}{|c|c|c|c|c|c|c|c|c|} \hline

&Total inputs&	1-5 min&	5-10 min&	10-15 min&	15-20 min&	20-40 min&	40-60 min& 	60-120 min\\ \hline

U1&	51990&	85&	7&	4&	1&	0&	0&	0\\ \hline
U2&	1094341&	86&	5&	3&	1&	0&	0&	0\\ \hline
U3&	104907&	45&	20&	17&	9&	5&	2&	0\\ \hline
U4&	124702&	49&	5&	0&	0&	0&	0&	0\\ \hline

\hline\end{tabular}
\end{table*}

Figure 8 shows the 4 users' input interval distribution within different time ranges.

\begin{figure}
\centering
\includegraphics[height=2.07in, width=3.5in]{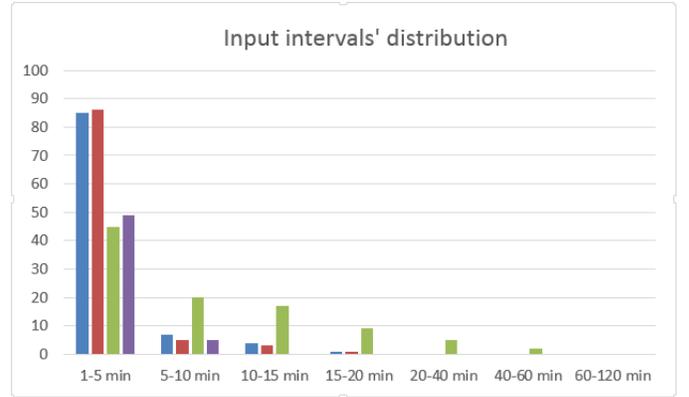}
\caption{the 4 users' input interval distribution within different time range.}
\end{figure}

\section{CONCLUSION}
In this paper, we take document files as an example, trying to exploit the user's experiences and other attributes to re-find a document. To alleviate the user's cognitive burden in the memory recollection process, we use a wizard to split a complicate question into small independent steps. Technically, this method is applicable to other file types too, such as multimedia files, in that case, some attributes should be abandoned, such as printing experience; some attributes should be added, such as duration etc.\\

People often encounter difficult re-finding tasks \cite{Blanc-Brude2007}. Some of them may spend a user hours of time \cite{Elsweiler2011}, or until the user gives up, what is worse, after several failed attempts, people may get frustrated or even mad. Therefore, it is necessary to develop new methods and new systems to deal with difficult re-find tasks. Our contributions can be concluded as follows:\\

1.	We devised a wizard based method to support difficult re-finding tasks.\\

2.	We propose to record all the possible attributes, including a document's intrinsic attributes and the user's experiences about the document, if any of them can get into the user's long term memory, the recorded attributes will be useful for re-finding.\\

3.	We devised an interface which can alleviate the user's cognitive burden when they are recollecting memories about a document.\\

4.	We devised an algorithm to calculate a user's precise reading time on a PDF document.\\

5.	We developed a PDF document re-finding system and evaluated its performance.

\section{DISSCUSSION}
People's reading and re-finding behaviors are very complicated and unpredictable, to develop a well performing file re-finding system, there are some other issues should be considered.

\subsection{Privacy issues}
Our re-finding system will inevitably record all the essential document data and the user's process experiences of the documents, these recorded data may involve the user's privacy.
Fortunately, the re-finding system mainly works for a user privately, it only runs in the user's personal computer, all the recorded data is stored in local disk, the start-up of the re-finding system is password protected. To be safer, we provide the function to delete the sensitive information optionally, such as deleting all the records of a particular document.

\subsection{Sequence of the questions}
To simplify the implementation, at present we assume each question is independent with others, the user's answer to one question has no relation with other questions. In practice, some attributes may have a connection with others. We plan to explore these connections and exploit it to improve re-finding in further works.

\subsection{What happens if the user's memory about an attribute is incorrect?}
People usually have vague memories for remote documents, it is likely that the user has an incorrect memory about a particular attribute, these incorrect information will lead to a failed re-finding. Further study will aim to find which set of attributes are inclined to be recollected wrongly.

%
\bibliographystyle{abbrv}
\bibliography{my}  
%
%

\balancecolumns 
\end{document}